\documentclass[sigconf,nonacm]{acmart}

\AtBeginDocument{%
  }

\setcopyright{none}
\copyrightyear{2026}
\acmYear{2026}

\acmConference[CIKM '26]{the 35th ACM International Conference on Information and Knowledge Management}{November 7--11, 2026}{Rome, Italy}
\acmDOI{}
\acmISBN{}

\startPage{1}

\usepackage{multirow}
\usepackage{xcolor}
\usepackage{booktabs}
\usepackage{graphicx}
\usepackage{enumitem}
\usepackage{algorithm}
\usepackage{algpseudocode}
\usepackage{subcaption}
\usepackage{tikz}
\usetikzlibrary{shapes.geometric,arrows.meta,positioning,fit,backgrounds,calc,decorations.pathreplacing}
\DeclareMathOperator*{\argmax}{argmax}

\begin{document}

\title{DREAM: Dynamic Refinement of Early Assignment Mappings}

\author{\texorpdfstring{Liwei Guan \quad Huanjie Wang \quad Hongwei Zhang \quad Linxun Chen\textsuperscript{\textdagger} \quad Zhaojie Liu}{Liwei Guan, Huanjie Wang, Hongwei Zhang, Linxun Chen, Zhaojie Liu}}
\affiliation{%
    \institution{Kuaishou Technology}
    \city{Beijing}
    \country{China}
}
\email{{guanliwei, wanghuanjie, zhanghongwei08, chenxi36, zhaotianxing}@kuaishou.com}

\renewcommand{\shortauthors}{Guan et al.}

\begin{abstract}
Generative recommendation advances item retrieval by reformulating it as
autoregressive generation of Semantic IDs (SIDs), compact token sequences
that encode item semantics. While SIDs offer a strong semantic prior,
current SID-based methods assign each item a single static identifier
through offline tokenization before sufficient user feedback is observed. For
cold-start items, this one-shot commitment produces poorly discriminative
codes, generating misaligned paths that remain unrefined because the
associated tokens are rarely sampled during training. We identify this early
static commitment, not model capacity, as the fundamental cold-start bottleneck
in SID-based generative recommendation. To overcome this bottleneck and bridge
the disjoint objectives of tokenization and generation, we propose \textbf{DREAM}
(\textbf{D}ynamic \textbf{R}efinement of \textbf{E}arly \textbf{A}ssignment
\textbf{M}appings), a three-stage framework that resolves this flaw through
progressive refinement. First, an intent-aware tokenizer rebuilds the SID space
through counterfactual contrastive learning, generating a diverse pool of
behavior-aligned candidates per cold-start item. Second, the frozen
recommendation backbone serves as an evaluator, selecting the most reliable
candidate based on multi-context user support without retraining. Third, a
dynamic beam mechanism maintains multiple weighted SID hypotheses throughout
training and inference, preventing premature collapse to a single assignment.
Extensive experiments on three Amazon benchmarks show that DREAM substantially
outperforms state-of-the-art generative and sequential baselines on cold-start
metrics.
\end{abstract}



\maketitle
\begingroup
\renewcommand{\thefootnote}{\textdagger}
\footnotetext{Corresponding author.}
\endgroup

\section{Introduction}
\label{sec:introduction}

Recommender systems have traditionally used cascaded recommender
architectures that optimize candidate retrieval and ranking
separately~\cite{youtubrednn,widedeep,deepfm,din}.
Generative Recommendation (GR) instead trains recommendation as
next-token prediction over item token
sequences~\cite{tiger,p5,dsi,nci,idgenrec}.
Each item is assigned a compact \emph{Semantic Item ID} (SID), a
discrete token sequence produced by hierarchical quantization
(e.g., RQ-VAE~\cite{rqvae} or RQ-KMeans~\cite{qarm}) of multimodal
item features~\cite{idvsmorec}; an autoregressive Transformer
backbone then generates the target item's SID from the user's
interaction history.
This formulation turns candidate generation into autoregressive
decoding over a structured identifier space, enables semantic transfer
through a shared token vocabulary, and has yielded strong results on
both academic benchmarks and industrial
deployments~\cite{hstu,onerec,onerecv2,grsurveyLi,grsurveyTri}.

Despite these advances, cold-start failure differs qualitatively
between SID-based generative recommendation and classical
ID-based recommenders.
In ID-based sequential models, a cold item usually remains in the
candidate set: it keeps an item embedding and receives a relevance
score, although that score may be unreliable when learned from few
interactions~\cite{maml,melu,clcrec,dropoutnet,heater,unisrec}.
In SID-based generative recommendation, the same sparsity can make the
item unreachable rather than merely weakly scored: the offline
tokenizer commits each item to one SID path before user feedback is
observed, and constrained decoding permits only registered paths.
When a cold-start item's assigned path is misaligned, the item is
sampled too rarely during training to receive corrective gradient, and
no alternative path is exposed to the decoder at inference.
Better upstream embeddings can improve the initial SID quality, yet
they do not by themselves remove this one-shot commitment.
We dub this bottleneck \emph{early static single-path commitment} and
decompose it into three coupled factors that DREAM targets.
The first factor is \emph{unsupported assignment}, where the SID is
chosen from content or pre-trained signals alone, without observed
interactions supporting its collaborative role.
The second is \emph{premature commitment}, where supervision is tied
to the single assigned path while cold-start gradient remains too
sparse to repair the choice.
The third is the \emph{inference-time single-path constraint}, where
constrained decoding registers only the assigned SID, so misaligned
cold-start items offer no alternative at test time.

Recent GR optimizations improve SID construction and align it with the
recommender, but they are largely cold-start-agnostic:
they target average tokenization or training quality rather than the
commitment decisions required for sparse items.
SID-based generative recommenders~\cite{lcrec,letter,tiger} treat
the SID as a fixed input produced offline and focus on backbone
training; cold behaviour is bounded by whatever the upstream
tokenizer commits.
Dedicated tokenizer designs improve offline quantization through
collaborative coupling~\cite{ccgen,eager}, learnable
codebooks~\cite{letter}, unified Semantic ID
representations~\cite{usid}, long parallel SIDs~\cite{rpg}, or
differentiable and drift-aware identifier learning~\cite{diger,dact}.
End-to-end co-evolution approaches~\cite{etorec,bloger,pit} offer
an elegant alternative by jointly optimizing the tokenizer and the
recommender in a single objective, enabling continuous mutual
alignment.
However, under cold-start conditions, joint objectives can still be
dominated by warm items because their interactions provide much denser
gradient signal, leaving cold SID assignments under-corrected.
Moreover, making discrete SID sampling differentiable usually requires
relaxations such as Gumbel-Softmax~\cite{gumbel}, which can be unstable
when cold-start items provide only sparse interaction supervision.
More importantly, these approaches still force each cold-start item
toward a single committed SID: they do not explicitly support deferring
the decision under insufficient interaction support, abstaining from
unreliable updates, or preserving multiple valid paths for inference.
No existing thread simultaneously (i) anchors cold SIDs to
prior-supported alternatives, (ii) defers commitment under
insufficient interaction support, and (iii) preserves multiple paths so
that the decoder can still reach the item at inference.
This motivates an explicit \emph{decoupling} strategy: rather than
asking one joint objective to both reconstruct the SID space and
commit each cold-start item to a path, DREAM separates prior construction,
support-gated commitment, and inference-time path preservation.

We instantiate this decoupling as \textbf{DREAM}, a progressive
refinement framework that drives a cold-start item's SID assignment
through prior-support repair, conservative commitment, and multi-path
recovery rather than three independent patches
(Figure~\ref{fig:framework}).
\textbf{Stage~1}, \emph{Collaborative-Aware Refined Tokenization}
(CART), performs \emph{prior-support repair}: it rebuilds the SID
space with collaborative geometry supervision and, for each cold-start item,
exposes a small prior-supported candidate pool of plausible SID paths,
turning subsequent stages into a well-bounded selection-and-preservation
problem.
\textbf{Stage~2}, \emph{User-Conditioned Candidate Condensation}
(UC3), performs \emph{conservative commitment}: the frozen
recommendation backbone serves as a zero-training-cost evaluator, and
a cold-start item's SID is updated only when confidence-weighted
multi-context votes pass explicit support and margin gates; otherwise,
the CART assignment is retained.
\textbf{Stage~3}, \emph{Cold-Preserved Dynamic Beam Evolution}
(CPDE), enables \emph{multi-path recovery}: it retains the surviving
SID alternatives, registers all of them in the constrained-decoding
trie, and performs multi-path inference; gradient isolation via
LoRA~\cite{hu2022lora} keeps the sparse cold updates from disrupting
warm-item training.
Each stage operates on the stabilized output of the previous one,
and conservative safeguards, including confidence gating in UC3 and
momentum-damped beam updates in CPDE, structurally bound error
propagation so that upstream imprecision does not compound.
On three Amazon benchmarks, DREAM achieves the best score on
\emph{all 18} cold-start metrics (improvements ranging from
$4\times$ to $12\times$ over the strongest baseline) while remaining
competitive on overall metrics (Section~\ref{sec:experiments}).

\noindent The main contributions of this paper are:
\begin{itemize}[nosep, leftmargin=*]
  \item \textbf{Problem formulation.} We identify early static
    single-path commitment, which induces path misalignment under the
    single-path assignment constraint, as a primary structural
    bottleneck of SID-based generative recommendation for cold-start
    items.
    We decompose it into three compounding factors (unsupported
    assignment, premature commitment, and the inference-time
    single-path constraint)
    and show that no existing method addresses all three jointly.

  \item \textbf{Conservative progressive refinement framework.}
    We propose DREAM, a three-stage system organized around
    prior-support repair, conservative commitment, and multi-path
    recovery.
    CART provides prior-supported SID candidates
    (Section~\ref{sec:cart}), UC3 applies confidence-weighted voting
    with explicit support and margin gates (Section~\ref{sec:uc3}),
    and CPDE preserves multi-path recovery through dynamic beam
    decoding with gradient isolation (Section~\ref{sec:cpde}).
    These conservative mechanisms bound error propagation across stages
    and protect warm-item performance during cold-start refinement.

  \item \textbf{Comprehensive empirical validation.} DREAM achieves
    the best result on all 18 cold-start metrics across three Amazon
    benchmarks, with improvements of roughly $4\times$ to $12\times$
    over the strongest baselines, while remaining competitive on
    overall metrics.
    Stage-wise mechanism diagnosis on Sports confirms each
    component's causal contribution.
\end{itemize}

\section{Related Work}
\label{sec:related_work}

\noindent\textbf{SID-Based Generative Recommendation and Item Tokenization.}
Generative retrieval first showed that retrieval targets can be generated as
identifiers: DSI~\cite{dsi} stores documents in a differentiable search index,
and NCI~\cite{nci} improves neural corpus indexing with document identifiers.
In recommendation, P5~\cite{p5} formulates recommendation with textual prompts
and item identifiers, while TIGER~\cite{tiger} introduces RQ-VAE semantic IDs
(SIDs) for generative recommendation.
LC-Rec~\cite{lcrec} and LETTER~\cite{letter} further adapt LLM-based
recommendation with collaborative semantics and representation alignment, and
large-scale systems such as HSTU~\cite{hstu}, OneRec and
OneRec-V2~\cite{onerec,onerecv2}, and QARM~\cite{qarm} demonstrate the
practical value of generative recommendation and compact item tokenization.

Recent work improves identifier construction and alignment from complementary
directions.
CCGen~\cite{ccgen} and EAGER~\cite{eager} inject collaborative or
behavior-semantic signals into tokenization; LETTER~\cite{letter},
USID~\cite{usid}, DiscRec~\cite{discrec}, and RPG~\cite{rpg} learn
more adaptive, unified, disentangled, or parallel item identifiers;
and ETEGRec~\cite{etorec}, BLOGER~\cite{bloger}, DIGER~\cite{diger},
DACT~\cite{dact}, and PIT~\cite{pit} show that token assignments can
be adapted with recommendation objectives, differentiable updates,
continual refinement, or personalized contexts.
DREAM is inspired by the broader idea that SID assignment need not remain
static, but it does not optimize the tokenizer and recommender in a single
end-to-end objective; instead, it isolates the cold-start commitment decision
after prior-supported candidates have been constructed.
Multi-view identifier methods such as MINDER~\cite{minder}, NOVO~\cite{novo},
Multi-DSI~\cite{multidsi}, and Pctx~\cite{pctx} further suggest that one target
may benefit from more than one valid identifier.
These methods improve identifier quality, alignment, or view diversity, but
most still commit each item to one primary path before cold-start interaction
support is available.
DREAM instead treats cold-start SID assignment as a staged commitment problem:
it constructs prior-supported candidates, commits only under sufficient
interaction support, and preserves multiple valid paths for constrained
decoding.

\noindent\textbf{Cold-Start Recommendation and Inductive Generative Retrieval.}
Classical cold-start methods mainly improve transferable representations while
the item remains in a global candidate pool: MeLU~\cite{melu} meta-learns
preference estimators, DropoutNet~\cite{dropoutnet} simulates missing
interactions, CLCRec~\cite{clcrec} uses contrastive alignment,
Heater~\cite{heater} applies randomized training with mixture-of-experts
transformation, and UniSRec~\cite{unisrec} learns universal sequence
representations for transfer.
Inductive generative retrieval addresses a harder setting in which new items
must be reached by a generator; for example, SpecGR~\cite{specgr} uses
speculative decoding to route unseen items in a plug-and-play generative
recommendation framework, and GenRecEdit~\cite{genrecedit} adapts model
editing to inject cold-start item knowledge without full retraining.
Because the offline SID assignment is still treated as an input prior,
inference-time speculation or model editing alone does not explicitly repair a
poorly supported cold-start path before commitment.
However, these methods improve continuous representations, inductive routing,
or model parameters rather than explicitly repairing the discrete SID
commitment made by the tokenizer, deferring commitment under insufficient
interaction support, and registering multiple cold-start paths in the decoding
trie.

\section{Methodology}
\label{sec:methodology}

\subsection{Preliminaries}
\label{sec:prelim}

\noindent\textbf{SID-Based Generative Recommendation.}
Let $\mathcal{U}$ and $\mathcal{I}$ denote the user and item sets.
For a training instance, $h_u=(i_1,\ldots,i_m)$ is the observed
interaction history of user $u\in\mathcal{U}$, and $i^*\in\mathcal{I}$
is the ground-truth next item.
Each item $i$ is associated with a content embedding
$\mathbf{e}_i \in \mathbb{R}^{d_e}$ and assigned a length-$L$
Semantic ID (SID).
An SID index is a mapping
$\Phi:\mathcal{I}\to\prod_{l=1}^{L}\mathcal{V}_l$, where
$\mathcal{V}_l$ is the token vocabulary at SID position $l$ and
$|\mathcal{V}_l|{=}V_l$.
Thus $\Phi(i)=s_i=[t_1^i,\ldots,t_L^i]$ with
$t_l^i\in\mathcal{V}_l$.
Following~\cite{tiger,lcrec}, the initial static index $\Phi^0$ is
produced offline by an RQ-VAE or RQ-KMeans quantizer~\cite{rqvae,qarm}
over these item embeddings.
Given an index $\Phi$, the history is represented in SID space as
$\mathbf{s}_{h_u}=(\Phi(i_1),\ldots,\Phi(i_m))$, and a generative
recommender $\mathcal{M}_\theta$ is trained to maximize the
autoregressive likelihood of the target SID $\Phi(i^*)$:
\begin{equation}
  \mathcal{L}_{\mathrm{rec}} = -\sum_{l=1}^{L}
    \log P_\theta\!\left(t_l^{i^*} \mid \mathbf{s}_{h_u},\,
      t_1^{i^*},\ldots,t_{l-1}^{i^*}\right).
  \label{eq:rec_loss}
\end{equation}
At inference, constrained decoding restricts generation to the
registered path set
$\mathcal{R}(\Phi)\!=\!\{\Phi(i)\mid i\in\mathcal{I}\}$, so $\Phi$
determines both the training targets and the reachable decoding
space.

\noindent\textbf{Problem Statement.}
Let $f_i$ be the number of training interactions containing item $i$.
Given a cold-start threshold $n_c$ (set to $5$ in our experiments), we
partition $\mathcal{I}$ into cold-start items
$\mathcal{I}_c = \{i \mid f_i \leq n_c\}$ and warm items
$\mathcal{I}_w = \mathcal{I} \setminus \mathcal{I}_c$.
Under $\Phi^0$ every cold-start item is committed to one SID before
sufficient interaction support is available, inducing the three
coupled risks of early static single-path commitment introduced in
\S\ref{sec:introduction}.
Given $\Phi^0$, item content embeddings $\{\mathbf{e}_i\}_{i\in\mathcal{I}}$,
and training interactions, our goal is to produce a refined
single-path index $\Phi_\mathrm{C}$ for backward-compatible decoding
and, for each cold-start item, a small alternative SID set
$\mathcal{B}_i$ for multi-path inference, while bounding warm-item
perturbation.

\subsection{Framework Overview}
\label{sec:overview}

\begin{figure*}[t]
\centering
\includegraphics[width=\textwidth]{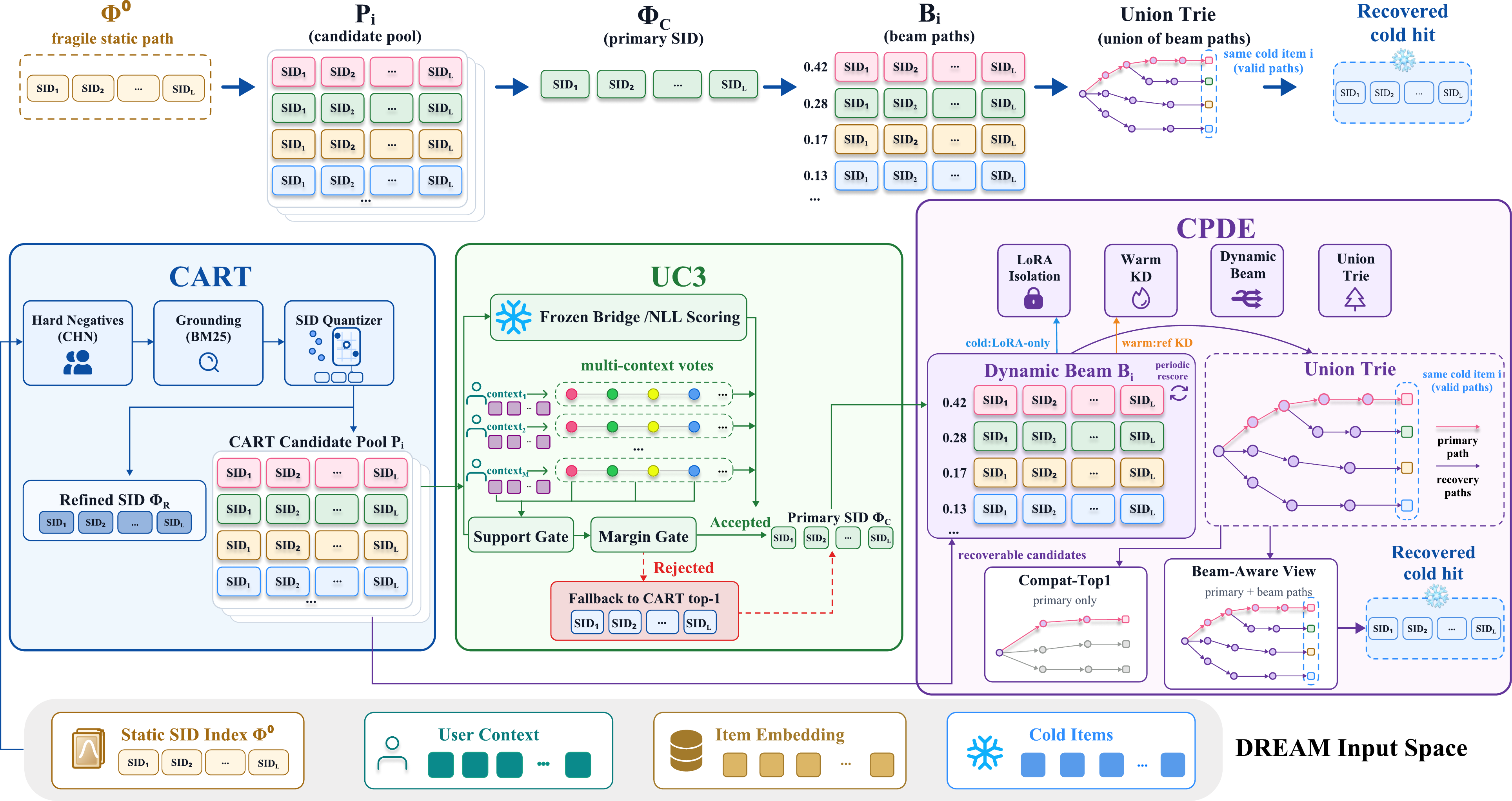}
\caption{Overview of the DREAM framework.
The top trajectory shows SID evolution from the fragile static path $\Phi^0$ to CART candidate pool $\mathcal{P}_i$, UC3 primary SID $\Phi_\mathrm{C}$, CPDE beam paths $\mathcal{B}_i$, and a union trie for cold-hit recovery.
CART injects collaborative signal into SID refinement, UC3 selects reliable candidates through multi-context votes and conservative gates, and CPDE performs beam-aware inference over valid trie paths.}
\Description{A wide DREAM architecture diagram. The top row shows semantic ID evolution from a fragile static SID path to a candidate pool, primary SID, beam paths, union trie, and recovered cold hit. The main row contains CART, UC3, and CPDE modules; the bottom row lists the DREAM input space.}
\label{fig:framework}
\end{figure*}

As illustrated in Figure~\ref{fig:framework}, DREAM refines the
initial index $\Phi^0$ through three stages.
\textbf{CART} (\S\ref{sec:cart}) rebuilds the SID space with
collaborative supervision and outputs a CART-refined index
$\Phi_\mathrm{R}$ together with a bounded prior-supported candidate
pool $\mathcal{P}_i$ for each cold-start item $i$.
\textbf{UC3} (\S\ref{sec:uc3}) treats the frozen bridge model as a
zero-training-cost evaluator, converts each pool into
confidence-weighted votes $\{v_i^{(k)}\}$, and commits to a
single-path index $\Phi_\mathrm{C}$ only when explicit support and
margin gates pass; otherwise it abstains at the CART assignment.
\textbf{CPDE} (\S\ref{sec:cpde}) preserves the surviving SID
alternatives as a dynamic beam $\mathcal{B}_i$, isolates cold
gradients via LoRA~\cite{hu2022lora}, and registers all retained paths in the
beam-aware constrained-decoding trie $\mathcal{T}_{\mathrm{BA}}$ for
multi-path inference.
To prevent upstream imprecision from compounding across stages, each
stage is decoupled by an explicit gating step: UC3's support and
margin thresholds filter low-support rewrites before commitment,
while CPDE updates beam weights through an exponential moving average
(EMA) update~\cite{tarvainen2017mean,moco} that suppresses noisy
single-step shifts
(\S\ref{sec:experiments} reports the resulting stage-wise stability).

\subsection{Collaborative-Aware Refined Tokenization}
\label{sec:cart}

CART repairs the prior support of cold-start SID paths before the recommender is
asked to make a final commitment.
Its role is not to arbitrarily rewrite identifiers, but to replace a
content-dominated top-1 assignment with a collaboration-aware local candidate
pool.
To make this prior useful under sparsity, CART combines collaborative item
representations, intent-discriminative hard negatives, and diversity-aware
quantization into a single tokenization objective.

\noindent\textbf{Item Representation.}
CART starts from the pre-computed item embedding $\mathbf{e}_i$ and
augments it with a learnable per-item collaborative residual
$\mathbf{r}_i \in \mathbb{R}^d$, initialized to zero:
\begin{equation}
  \mathbf{z}_i = \mathrm{LN}\!\left(W_c\,\mathbf{e}_i + \mathbf{r}_i\right),
  \label{eq:item_repr}
\end{equation}
where $W_c$ is a linear content projection and $\mathrm{LN}$ is layer normalization.
The zero initialization keeps early training anchored in content geometry
while $\mathbf{r}_i$ gradually accumulates interaction-derived structure.
Even for cold-start items, $\mathbf{r}_i$ receives gradient from two complementary
paths: collaborative geometry via $\mathcal{L}_\mathrm{cal}$ and
counterfactual hard-negative contrastive signal via
$\mathcal{L}_\mathrm{NCE}$ (detailed below), so the representation
remains discriminative under interaction sparsity.

\noindent\textbf{Counterfactual Hard Negative Mining (CHN).}
Random negatives are too easy to teach intent-level discrimination, so
we mine intent-discriminative hard negatives in two steps.

\emph{Step 1: Counterfactual generation.}
For each item $i$, we prompt an LLM with a domain-specific template
that keeps the product family unchanged but flips the core purchase
intent along a semantically salient axis
(e.g., hydrating $\to$ oil-control; trail-running $\to$ road-running).
A rule-based fallback handles generation failures to maintain full coverage.

\emph{Step 2: Catalog grounding via sparse retrieval.}
The generated description is synthetic, so using it directly would
introduce out-of-distribution embeddings.
We instead use it as a query for BM25~\cite{bm25}, a standard sparse
retrieval method, to retrieve the $K$ most lexically matched real
items from the catalog:
\begin{equation}
  \mathcal{N}_i = \mathrm{BM25\text{-}top}\!\left(
    \tilde{\mathbf{x}}_i,\; \mathcal{I} \setminus \{i\},\; K\right).
  \label{eq:bm25}
\end{equation}
We choose BM25 because lexical overlap is the dominant source of
confusability at the SID level.
We apply CHN to all items: warm items supply dense comparative
context that calibrates the geometric boundary around cold-start
items, while cold-start items receive hard-negative signal even when
interactions are sparse.

\noindent\textbf{SID Quantization with Diversity Regularization.}
We tokenize each item into an $L$-position SID using codebooks
$\mathbf{C}_l \in \mathbb{R}^{V_l \times d}$.
A projection $f_q$ maps $\mathbf{z}_i$ to per-position queries $\mathbf{q}_{i,l}$,
and the assignment logit is computed by cosine similarity scaled by a
learnable temperature $\tau_s$:
\begin{equation}
  \ell_{i,l,k} = \tau_s \cdot
    \frac{\mathbf{q}_{i,l}^\top \mathbf{c}_{l,k}}
         {\|\mathbf{q}_{i,l}\|\,\|\mathbf{c}_{l,k}\|}.
  \label{eq:logits}
\end{equation}
Cosine normalization prevents high-magnitude entries from dominating
irrespective of semantic fit.
We train with Gumbel-Softmax sampling~\cite{gumbel,gumbel_maddison} under
an exponentially annealed temperature
$\tau(t) = \tau_0 \cdot (\tau_\mathrm{end}/\tau_0)^{t/T_\mathrm{max}}$,
paired with a Straight-Through Estimator
(STE)~\cite{ste_bengio} so that the discrete one-hot assignment is used in the
forward pass while the continuous soft assignment provides the backward
gradient.

To keep the candidate prior from collapsing into a small set of overused codes,
we add a diversity regularizer that penalizes concentration in the batch-mean
assignment distribution:
\begin{equation}
  \mathcal{L}_{\mathrm{div}} = \frac{1}{L}\sum_{l=1}^{L}
    \Bigl(1 - \tfrac{H(\bar{\mathbf{a}}_l)}{\log V_l}\Bigr),
  \label{eq:div_loss}
\end{equation}
where $\bar{\mathbf{a}}_l = \frac{1}{|\mathcal{B}|}\sum_{i}\mathrm{softmax}(\ell_{i,l})$
and $H(\cdot)$ is Shannon entropy.
$\mathcal{L}_{\mathrm{div}}$ is a soft penalty weighted by
$\lambda_\mathrm{div}$ that suppresses batch-level code collapse,
while the anchor loss introduced below preserves semantically
meaningful token concentration.

\noindent\textbf{Training Objectives.}
CART trains the item representations and quantizer with four terms.

\noindent\emph{Contrastive (NCE).}
A momentum encoder, updated as an EMA of $W_c$ and $\{\mathbf{r}_i\}$ with
coefficient $m$ following standard momentum-encoder practice~\cite{moco},
produces consistent positive keys $\hat{\mathbf{z}}_i^+$.
Combined with a FIFO negative queue $\mathcal{Q}$ and the CHN pool $\mathcal{N}_i$,
the position-averaged InfoNCE loss is:
\begin{equation}
  \mathcal{L}_{\mathrm{NCE}} = -\frac{1}{L}\sum_{l=1}^{L}\log
    \frac{e^{\hat{\mathbf{s}}_{i,l}^\top\hat{\mathbf{z}}_i^+/\tau_n}}
         {e^{\hat{\mathbf{s}}_{i,l}^\top\hat{\mathbf{z}}_i^+/\tau_n}
          + \sum_{j\in\mathcal{Q}\cup\mathcal{N}_i}
            e^{\hat{\mathbf{s}}_{i,l}^\top\hat{\mathbf{z}}_j^+/\tau_n}},
  \label{eq:nce}
\end{equation}
where $\hat{\mathbf{s}}_{i,l}$ is the normalized STE token embedding.

\noindent\emph{Collaborative alignment (CAL).}
To inject collaborative structure that contrastive learning alone
does not capture, we add a target-conditioned CTR task.
For each history item $j$, we construct the interaction feature
$\mathbf{f}_j = [\mathbf{z}_{i_j};\mathbf{z}_{i^*};\mathbf{z}_{i_j}{-}\mathbf{z}_{i^*};
\mathbf{z}_{i_j}{\odot}\mathbf{z}_{i^*}] \in \mathbb{R}^{4d}$
and compute attention weights via a two-layer MLP
$a\!:\mathbb{R}^{4d}\!\to\!\mathbb{R}$:
\begin{equation}
  \alpha_j = \frac{\exp\!\bigl(\mathbf{v}_a^\top\mathrm{ReLU}(W_a\mathbf{f}_j + \mathbf{b}_a)\bigr)}
    {\sum_{j'}\exp\!\bigl(\mathbf{v}_a^\top\mathrm{ReLU}(W_a\mathbf{f}_{j'} + \mathbf{b}_a)\bigr)},
  \quad
  \mathbf{u} = \textstyle\sum_j\alpha_j\,\mathrm{sg}(\mathbf{z}_{i_j}),
  \label{eq:user_repr}
\end{equation}
where $W_a\!\in\!\mathbb{R}^{d_h\times 4d}$,
$\mathbf{v}_a\!\in\!\mathbb{R}^{d_h}$, and $d_h$ is the hidden size.
A prediction head $g\!:\mathbb{R}^{3d}\!\to\!\mathbb{R}$ (same architecture)
takes $[\mathbf{u};\mathbf{z};\mathbf{u}{\odot}\mathbf{z}]$ and outputs a
logit for the BCE objective:
\begin{equation}
\begin{aligned}
  \mathcal{L}_{\mathrm{cal}} =
    &\mathrm{BCE}\!\bigl(\sigma(g([\mathbf{u};\mathbf{z}_{i^*};
      \mathbf{u}{\odot}\mathbf{z}_{i^*}])),\,1\bigr) \\
    &+ \mathrm{BCE}\!\bigl(\sigma(g([\mathbf{u};\mathbf{z}_{j^-};
      \mathbf{u}{\odot}\mathbf{z}_{j^-}])),\,0\bigr).
\end{aligned}
  \label{eq:cal}
\end{equation}
The stop-gradient on each history embedding $\mathbf{z}_{i_j}$
prevents the CTR task from modifying history representations through
co-occurrence shortcuts, while the target $\mathbf{z}_{i^*}$ still
receives full gradient from $\mathcal{L}_\mathrm{NCE}$ and
$\mathcal{L}_\mathrm{cal}$.

\noindent\emph{Static anchor.}
To keep the refined SID space compatible with the downstream token vocabulary,
we regularize toward the existing static assignment $s_i^0$:
\begin{equation}
  \mathcal{L}_{\mathrm{anc}} = \frac{1}{L}\sum_{l=1}^{L}
    \mathrm{CE}(\ell_{i,l},\; t_l^{0,i}).
  \label{eq:anchor}
\end{equation}

The full CART objective is:
\begin{equation}
  \mathcal{L}_{\mathrm{CART}} =
    \mathcal{L}_{\mathrm{NCE}} + \mathcal{L}_{\mathrm{cal}}
    + \lambda_{\mathrm{div}}\,\mathcal{L}_{\mathrm{div}}
    + \lambda_{\mathrm{anc}}\,\mathcal{L}_{\mathrm{anc}},
  \label{eq:cart_total}
\end{equation}
where $\mathcal{L}_{\mathrm{NCE}}$ and $\mathcal{L}_{\mathrm{cal}}$ are
primary objectives with unit weight, and $\lambda_{\mathrm{div}}$,
$\lambda_{\mathrm{anc}}$ are hyperparameters controlling regularization strength.
Because $\mathcal{L}_\mathrm{anc}$ constrains the refined SIDs to stay close
to the original static index, warm items undergo only minor adjustments rather
than wholesale reassignment; the downstream adaptation cost is limited to a
single bridge fine-tuning pass on the updated index, which is lightweight
relative to the cold-start gains it enables.

After training, CART scores a candidate SID
$s=[t_1,\ldots,t_L]$ by the average token log-probability induced by the
learned quantizer:
\begin{equation}
  q_{\mathrm{CART}}(s\mid i)
  = \frac{1}{L}\sum_{l=1}^{L}
    \log \mathrm{softmax}(\ell_{i,l})_{t_l}.
  \label{eq:cart_path_score}
\end{equation}
For each item, we decode the $K$ highest-scoring SID sequences from the learned
quantizer.
The top-ranked sequence becomes the new primary assignment (the refined index),
and the full top-$K$ set forms a bounded candidate pool that the next stage
(UC3) will evaluate under observed user support:
\begin{equation}
  \Phi_\mathrm{R}(i)=s_i^{(1)},\qquad
  \mathcal{P}_i=\{(s_i^{(k)},p_i^{(k)})\}_{k=1}^{K}
  = \mathrm{TopK}_{s}\,q_{\mathrm{CART}}(s\mid i),
  \label{eq:cart_prior}
\end{equation}
where $p_i^{(k)}$ denotes the corresponding CART path score.
For warm items, only $\Phi_\mathrm{R}$ is used as the updated single-path
index; for cold-start items, the full candidate pool $\mathcal{P}_i$ is
passed to the downstream commitment and multi-path recovery stages.

\subsection{User-Conditioned Candidate Condensation}
\label{sec:uc3}

CART equips each cold-start item with a prior-supported pool $\mathcal{P}_i$ of $K$
ranked SID candidates, but the pool itself should not force an immediate
commitment.
The missing support signal is how each candidate behaves when generated under
real user histories.
UC3 obtains this support signal from the frozen bridge model:
for each candidate SID, it computes the teacher-forced negative log-likelihood
(NLL), which measures how likely the model is to generate that candidate
token-by-token given the user's interaction history.
This requires only $K$ forward passes per cold-start interaction with no
parameter updates.

Concretely, for each training interaction $(u, h_u, i)$ with $i\in\mathcal{I}_c$,
UC3 evaluates every candidate
$s^{(k)}_i \in \mathcal{P}_i$ under the frozen model:
\begin{equation}
  \mathrm{NLL}(s^{(k)}_i \mid u, h_u)
  = -\frac{1}{L}\sum_{l=1}^{L}
    \log P_\mathrm{frozen}\!\left(t^{(k)}_{i,l} \mid
      \mathbf{s}_{h_u}, t^{(k)}_{i,<l}\right).
  \label{eq:nll}
\end{equation}
UC3 first converts these losses into a per-context preference distribution:
\begin{equation}
  p_{u,i}^{(k)} =
  \frac{e^{-\mathrm{NLL}(s^{(k)}_i \mid u, h_u)/\tau_c}}
       {\sum_{k'} e^{-\mathrm{NLL}(s^{(k')}_i \mid u, h_u)/\tau_c}}.
  \label{eq:uc3_context_prob}
\end{equation}
Here $\tau_c$ is a fixed candidate-selection temperature.
Intuitively, each user context casts a vote over the candidate SIDs.
However, not every vote should count equally: when the frozen model assigns
nearly uniform probability to all candidates, it is telling us that this
context is not decisive.
We therefore weight each vote by an entropy-based confidence:
\begin{equation}
  \omega_{u,i}
  = 1 - \frac{H(\mathbf{p}_{u,i})}{\log K},
  \qquad
  H(\mathbf{p}_{u,i}) = -\sum_{k=1}^{K} p_{u,i}^{(k)}\log p_{u,i}^{(k)}.
  \label{eq:uc3_confidence}
\end{equation}
The final candidate score is a confidence-weighted consensus:
\begin{equation}
  v_i^{(k)} =
  \frac{\sum_{(u,h_u)\in\mathcal{D}_i}
    \omega_{u,i}\,p_{u,i}^{(k)}}
       {\sum_{(u,h_u)\in\mathcal{D}_i}\omega_{u,i} + \epsilon}.
  \label{eq:vote}
\end{equation}
where $\epsilon$ is a small numerical stabilizer.
Thus, contexts where the model is confident have larger influence, while
ambiguous contexts contribute little.
A candidate that scores well across heterogeneous, high-confidence user
contexts reflects a SID that is broadly compatible with the cold-start item's
diverse usage patterns, whereas a candidate that dominates only under a few
specific contexts receives lower aggregate support.

UC3 converts the vote scores into a commitment only when the support is both
sufficient and decisive.
Let $k^*=\argmax_k v_i^{(k)}$, and let $v_i^{[1]}$ and $v_i^{[2]}$ denote the
largest and second-largest vote scores.
The refined index is
\begin{equation}
  \Phi_\mathrm{C}(i)=
  \begin{cases}
    s_i^{(k^*)},
      & |\mathcal{D}_i|\geq\eta_\mathrm{sup}
        \;\land\; v_i^{[1]}-v_i^{[2]}\geq\eta_\mathrm{mar},\\
    \Phi_\mathrm{R}(i), & \mathrm{otherwise},
  \end{cases}
  \label{eq:uc3_commit}
\end{equation}
for cold-start items, while warm items keep $\Phi_\mathrm{C}(i)=\Phi_\mathrm{R}(i)$.
The fallback case is intentional abstention under weak support: UC3 updates a
cold-start item only when multi-context votes consistently improve upon the CART
prior, and otherwise retains the safer CART top-1 assignment.
The resulting index $\Phi_\mathrm{C}$ and the per-candidate vote scores
$\{v_i^{(k)}\}$ are passed to CPDE.

\subsection{Cold-Preserved Dynamic Beam Evolution}
\label{sec:cpde}

Even after UC3 selects a stable single path, cold-start relevance may retain
residual multi-path ambiguity.
Different user contexts can support different plausible SIDs from the CART
prior, and collapsing all of them into one top-1 path can remove recoverable
routes from the constrained-decoding trie.
At the same time, unrestricted fine-tuning is undesirable: sparse cold-start
gradients carry high variance and can perturb warm-item generation, while dense
warm gradients can dilute cold-specific adaptation.
CPDE therefore treats $\Phi_\mathrm{C}$ as the stable commitment point and
preserves a small set of residual alternatives for training and decoding.

To prevent sparse cold-start gradients from disrupting warm-item generation
while still allowing cold-specific adaptation, CPDE employs
\emph{cold-preserved gradient isolation}.
We inject rank-$r$ LoRA adapters~\cite{hu2022lora} into the query and value
projections of every attention layer.
When the model processes a cold-start sample, the entire backbone
(including all non-LoRA parameters) is frozen and its contribution is
stop-gradiented, so that gradients flow exclusively through the low-rank
adapter matrices:
\begin{equation}
  \mathbf{o}_\mathrm{cold} = \mathrm{sg}(W_0\mathbf{x})
    + \tfrac{\alpha_r}{r}\,\mathbf{B}\mathbf{A}\mathbf{x},
  \label{eq:cold_lora}
\end{equation}
where $\mathrm{sg}(\cdot)$ denotes stop-gradient.
For warm-item samples, both the backbone weights $W_0$ and the adapter receive
full gradient flow, preserving the model's warm-item generation ability.

For each cold-start item, CPDE maintains a dynamic beam
$\mathcal{B}_i = \{(s_i^{(k)}, w_i^{(k)})\}_{k=1}^B$ of $B$ weighted SID
hypotheses, initialised from $\mathcal{P}_i$ and $\Phi_\mathrm{C}$.
The model is trained with a soft multi-target objective weighted by the current
beam distribution $\tilde{w}_i^{(k)} \propto \exp(w_i^{(k)}/T)$:
\begin{equation}
  \mathcal{L}_\mathrm{soft}
  = -\sum_{k=1}^{B}\tilde{w}_i^{(k)}\cdot
    \frac{1}{L}\sum_{l=1}^{L}
    \log p_\theta\!\left(t_{i,l}^{(k)} \mid \mathbf{s}_{h_u}, t_{i,<l}^{(k)}\right).
  \label{eq:soft_ce}
\end{equation}
For warm items, generation ability is preserved via knowledge distillation from
the frozen reference model (a copy of the UC3 bridge checkpoint),
gated by the teacher's own output entropy to suppress distillation on
uncertain samples.
The full training objective is:
\begin{equation}
  \mathcal{L}_\mathrm{CPDE} = \lambda_s\,\mathcal{L}_\mathrm{soft}
    + \lambda_c\,\mathcal{L}_\mathrm{KL}^c
    + \lambda_w\,\mathcal{L}_\mathrm{KD},
  \label{eq:cpde_total}
\end{equation}
where $\mathcal{L}_\mathrm{KL}^c = \mathrm{KL}(p_\theta \| p_\mathrm{ref})$
is an asymmetric KL regulariser on SID-position logits that anchors the
cold-start path to the reference model, and $\mathcal{L}_\mathrm{KD}$ is
the entropy-gated warm-item distillation loss.

After a short warm-up of $E_0$ epochs, CPDE periodically refreshes the beam
weights every $K_e$ optimization steps using the current student model.
For each cold-start item observed in the recent optimization window
$\mathcal{W}_i$, the score of candidate $s_i^{(k)}$ is the average negative
teacher-forced NLL:
\begin{equation}
  \mathrm{score}_i^{(k)}
  = -\frac{1}{|\mathcal{W}_i|}
    \sum_{(u,h_u,i)\in\mathcal{W}_i}
    \mathrm{NLL}_\theta(s_i^{(k)}\mid u,h_u).
  \label{eq:beam_score}
\end{equation}
CPDE then updates beam weights via momentum EMA~\cite{tarvainen2017mean}:
\begin{equation}
  w_i^{(k)} \leftarrow \gamma_m\, w_i^{(k)}
    + (1 - \gamma_m)(\alpha_b + \mathrm{score}_i^{(k)}),
  \label{eq:beam_update}
\end{equation}
gated by a per-item confidence measure
$\mathrm{conf}(i) = \sigma(-\mathrm{NLL}_\mathrm{best}(i) + \delta)$,
where $\delta$ is a threshold and $\sigma$ is the sigmoid function.
When $\mathrm{conf}(i)$ is low, the update is withheld so that the beam
remains stable until the model accumulates stronger support.
This design creates a mutual refinement loop: as the model improves at
cold-start generation, beam weights are periodically refreshed to provide
cleaner training targets, which in turn accelerates model adaptation.
The initial warm-up period ($E_0$ epochs) and the confidence gate together
prevent premature beam collapse in early training.

At deployment, the LoRA weights are merged into the backbone with no
additional model-computation cost, and all $B$ retained beam paths per cold-start
item are pre-registered in the constrained-decoding trie to enable
multi-path inference (\S\ref{sec:inference}).
Algorithm~\ref{alg:dream} summarizes the full DREAM pipeline.

\begin{algorithm}[t]
\caption{DREAM: Dynamic Assignment Refinement and Multi-Path Inference}
\label{alg:dream}
\small
\begin{algorithmic}[1]
\Require Static index $\Phi^0$, item embeddings $\{\mathbf{e}_i\}$,
interaction data $\mathcal{D}$, cold-start set $\mathcal{I}_c$

\Statex \textbf{\textsc{Stage 1: CART} --- Prior-Support Repair (\S\ref{sec:cart})}
\State Build CHN pools $\mathcal{N}_i$ via BM25-grounded retrieval (Eq.~\ref{eq:bm25})
\State Train on all items with $\mathcal{L}_{\mathrm{CART}}$ (Eq.~\ref{eq:cart_total})
\State Decode refined index $\Phi_\mathrm{R}$; export top-$K$ prior $\mathcal{P}_i$ for $i\in\mathcal{I}_c$ (Eq.~\ref{eq:cart_prior})
\State Fine-tune bridge model $\mathcal{M}_\mathrm{R}$ on $\Phi_\mathrm{R}$

\Statex \textbf{\textsc{Stage 2: UC3} --- Conservative Commitment (\S\ref{sec:uc3})}
\For{each $i\in\mathcal{I}_c$}
  \State Score candidates via frozen-bridge NLL (Eq.~\ref{eq:nll})
  \State Aggregate confidence-weighted votes $v_i^{(k)}$ (Eq.~\ref{eq:vote})
  \State Commit $\Phi_\mathrm{C}(i)$ if support \& margin gates pass (Eq.~\ref{eq:uc3_commit}); else keep $\Phi_\mathrm{R}(i)$
\EndFor
\State Set $\Phi_\mathrm{C}(i)\!\leftarrow\!\Phi_\mathrm{R}(i)$ for $i\notin\mathcal{I}_c$; fine-tune $\mathcal{M}_\mathrm{C}$ on $\Phi_\mathrm{C}$

\Statex \textbf{\textsc{Stage 3: CPDE} --- Multi-Path Recovery (\S\ref{sec:cpde})}
\State Init beam $\mathcal{B}_i$ from $\Phi_\mathrm{C}(i)$ and UC3-weighted $\mathcal{P}_i$
\For{each training step $t$}
  \State Cold samples: LoRA-isolated gradient (Eq.~\ref{eq:cold_lora}), soft beam loss (Eq.~\ref{eq:soft_ce})
  \State Warm samples: entropy-gated distillation from frozen reference
  \If{past warm-up \textbf{and} $t\bmod K_e=0$}
    \State Rescore beams (Eq.~\ref{eq:beam_score}); momentum-update weights if confident (Eq.~\ref{eq:beam_update})
  \EndIf
\EndFor

\Statex \textbf{\textsc{Deploy:} Multi-Path Inference (\S\ref{sec:inference})}
\State Merge LoRA into backbone; build $\mathcal{T}_{\mathrm{C1}}$ (Eq.~\ref{eq:compat_trie}) and $\mathcal{T}_{\mathrm{BA}}$ (Eq.~\ref{eq:beam_aware_trie})
\State \Return $\mathcal{M}_{\mathrm{DREAM}},\;\Phi_\mathrm{C},\;\{\mathcal{B}_i\},\;\mathcal{T}_{\mathrm{C1}},\;\mathcal{T}_{\mathrm{BA}}$
\end{algorithmic}
\end{algorithm}

\subsection{Inference}
\label{sec:inference}

For cold-start items, DREAM performs multi-path inference as its primary
retrieval mechanism: all surviving beam SIDs are registered in the
constrained-decoding trie so that the model can reach a cold-start item
through any of its learned paths, increasing retrieval probability without
additional scoring modules.

At inference time, the merged model autoregressively generates the next item's
SID token by token:
\begin{equation}
  \hat{t}_l = \argmax_{v \in \mathcal{V}_l}
    P_\theta\!\left(v \mid \hat{t}_{<l},\, \mathbf{s}_{h_u}\right),
  \label{eq:inference}
\end{equation}
where $\hat{t}_{<l}$ denotes the previously generated prefix.
Constrained generation via a prefix trie~\cite{tiger,lcrec} restricts each
decoding step to valid continuations of registered SIDs.
DREAM instantiates two trie configurations from the same CPDE state.
Let $s_i^\dagger = \argmax_{k \in \{1,\dots,B\}} w_i^{(k)}$ denote the
highest-weight beam path for cold-start item $i$.
\textbf{Compat-Top1} keeps one path per item for backward-compatible
comparison with single-index baselines:
\begin{equation}
  \mathcal{T}_{\mathrm{C1}}
  =
  \{\Phi_\mathrm{C}(i)\mid i\in\mathcal{I}_w\}
  \cup
  \{s_i^\dagger\mid i\in\mathcal{I}_c\}.
  \label{eq:compat_trie}
\end{equation}
\textbf{Beam-Aware} registers all retained beam paths for each cold-start
item, enabling multi-path retrieval:
\begin{equation}
  \mathcal{T}_{\mathrm{BA}}
  =
  \{\Phi_\mathrm{C}(i)\mid i\in\mathcal{I}_w\}
  \cup
  \bigcup_{i\in\mathcal{I}_c}
    \{s_i^{(k)}\mid (s_i^{(k)},w_i^{(k)})\in\mathcal{B}_i\}.
  \label{eq:beam_aware_trie}
\end{equation}
A generated SID is resolved to the target if it matches any of that item's
registered paths.
Beam-Aware is reported as DREAM's primary view, since it reflects the
intended multi-path operational mode for cold-start items; Compat-Top1
is used in ablations to isolate single-path quality from multi-path
recovery.

\section{Experiments}
\label{sec:experiments}

We evaluate DREAM on three benchmark datasets and compare it against
ten competitive baselines. Our experiments address the following research
questions:
\textbf{(RQ1)} How effective is DREAM in cold-start item recommendation?
\textbf{(RQ2)} Do these cold-start gains translate into competitive
overall utility?
\textbf{(RQ3)} How do CART, UC3, and CPDE contribute to the final gains?
\textbf{(RQ4)} What is the trade-off between cold-start gains and
warm-item preservation?

\subsection{Experimental Setup}
\label{sec:setup}

\noindent\textbf{Datasets.}
We conduct experiments on three Amazon product review
categories: Beauty, Sports and Outdoors (Sports), and Toys and Games
(Toys).
We apply iterative core-5 filtering for users and core-3 filtering
for items.
For fair comparison, we uniformly employ a frozen
Llama-3-8B~\cite{llama3} to extract item text embeddings for initial
codebook construction in all SID-based methods (TIGER, LETTER, LC-Rec,
SpecGR, and DREAM), with a shared RQ-VAE~\cite{rqvae} configuration of
$L{=}4$ codebooks and $V_l{=}256$ codes per position.
For TIGER and LETTER, we adopt the implementation provided by
LETTER~\cite{letter}; for LC-Rec and DREAM, the backbone is fine-tuned
using LLaMA-7B~\cite{llama7b}, consistent with the original LC-Rec
design.
An item is classified as \emph{cold} if its total interaction count
$f_i \leq 5$ and as \emph{warm} otherwise.
Table~\ref{tab:dataset} summarizes the dataset statistics.
All three datasets exhibit extreme sparsity (interaction density below
$0.04\%$) and a substantial cold-start ratio ($19$--$24\%$ of test targets),
making them well suited for evaluating cold-start methods.

\begin{table}[t]
\centering
\normalsize
\setlength{\tabcolsep}{3.5pt}
\renewcommand{\arraystretch}{1.05}
\caption{Dataset statistics after iterative core filtering.
\emph{Cold\%}: fraction of test targets with $f_i \leq 5$.}
\label{tab:dataset}
\begin{tabular}{lrrrrc}
\toprule
Dataset & \#Users & \#Items & \#Inters & Avg.Len & Cold\% \\
\midrule
Beauty  & 32,106 & 26,595 & 283,907 & 8.8 & 20.6\% \\
Sports  & 49,133 & 37,602 & 412,439 & 8.4 & 19.1\% \\
Toys    & 31,261 & 29,840 & 271,242 & 8.7 & 23.8\% \\
\bottomrule
\end{tabular}
\end{table}

\noindent\textbf{Baselines.}
We compare DREAM against ten methods from three categories:
(1) \emph{ID-based sequential models}, including
\textbf{GRU4Rec}~\cite{gru4rec}, a GRU-based session recommender;
\textbf{SASRec}~\cite{sasrec}, a unidirectional self-attention model;
\textbf{BERT4Rec}~\cite{bert4rec}, a bidirectional masked Transformer;
\textbf{HGN}~\cite{hgn}, a hierarchical gating network; and
\textbf{HSTU}~\cite{hstu}, a large-scale sequential transducer;
(2) \emph{reasoning augmentation}, represented by
\textbf{ReaRec}~\cite{rearec}, which augments sequential recommenders
with inference-time multi-step reasoning; and
(3) \emph{SID-based generative retrieval}, including
\textbf{TIGER}~\cite{tiger}, the seminal RQ-VAE SID generative
retriever; \textbf{LETTER}~\cite{letter}, which augments TIGER with
alignment-based data augmentation and multi-task training;
\textbf{LC-Rec}~\cite{lcrec}, which integrates collaborative signals
into the SID-based LLM fine-tuning pipeline; and
\textbf{SpecGR}~\cite{specgr}, a plug-and-play inductive generative
recommendation framework that uses a drafter--verifier design to
propose and verify candidate items, including unseen items.

\noindent\textbf{Evaluation Protocol.}
We follow the standard leave-one-out protocol: for each user, the
last interaction is used for testing, the second-to-last for
validation, and the rest for training.
We report \textbf{Recall@$K$} (R@$K$) and \textbf{NDCG@$K$} (N@$K$)
at $K \in \{5, 10, 50\}$ with full-ranking over the entire item
catalog, computed separately for \emph{overall}, \emph{cold}, and
\emph{warm} subsets.
For DREAM we report the Beam-Aware view (\S\ref{sec:cpde}),
which evaluates against the multi-path union trie containing all
registered beam SIDs for each cold-start item and reduces to standard
single-index evaluation for single-path methods; the
backward-compatible Compat-Top1 view is reported in the
ablation study to isolate the contribution of multi-path decoding.

\noindent\textbf{Implementation.}
DREAM follows the LC-Rec~\cite{lcrec} LLaMA-7B setup, and all
SID-based methods use the shared $L{=}4$, $V_l{=}256$ configuration.
\emph{CART (Stage~1)} trains the tokenizer for 12 epochs (batch 64,
AdamW lr $5{\times}10^{-4}$) with NCE, collaborative alignment,
diversity ($\lambda_\mathrm{div}{=}0.1$), and static-anchor
($\lambda_\mathrm{anc}{=}0.5$) losses; it uses an $8{,}192$-entry FIFO
queue, $K{=}8$ counterfactual hard negatives, a $0.5$ warm
augmentation ratio, and 4 warm negatives per item, with
counterfactual descriptions generated by Llama-3.1-8B~\cite{llama3}
and BM25-grounded to the catalog.
After CART and UC3 index replacement, the backbone is fine-tuned for
12 epochs at lr $2{\times}10^{-5}$ with a cosine schedule on 8 GPUs
under DeepSpeed ZeRO-2.
\emph{UC3 (Stage~2)} scores the $K{=}8$ CART candidates with the
frozen bridge at $\tau_c{=}1.0$ and changes the SID only if the support
($\eta_\mathrm{sup}{=}3$) and margin ($\eta_\mathrm{mar}{=}0.05$)
gates both pass.
\emph{CPDE (Stage~3)} trains for 6 epochs at lr $2{\times}10^{-5}$
with rank-4 LoRA on attention $Q/V$, a beam width $B{=}4$, and beam
refresh every $K_e{=}50$ steps after $E_0{=}2$ warm-up epochs
($\gamma_m{=}0.7$).

\subsection{Cold-Start Performance}
\label{sec:cold_results}
\label{sec:main_results}

\begin{table*}[t]
\centering
\caption{Cold-start item recommendation performance.
DREAM achieves the best result across \textbf{all 18 cold-start metrics}
on three datasets, with $4.3\times$ to $11.5\times$ improvements over the
strongest per-metric baseline (peak: $11.5\times$ N@50 on Sports).}
\label{tab:cold}
\resizebox{\textwidth}{!}{
\begin{tabular}{l|cccccc|cccccc|cccccc}
\toprule
\multirow{2}{*}{Method} & \multicolumn{6}{c|}{Beauty} & \multicolumn{6}{c|}{Sports} & \multicolumn{6}{c}{Toys} \\
 & R@5 & N@5 & R@10 & N@10 & R@50 & N@50 & R@5 & N@5 & R@10 & N@10 & R@50 & N@50 & R@5 & N@5 & R@10 & N@10 & R@50 & N@50 \\
\midrule
GRU4Rec & 0.0015 & 0.0008 & 0.0027 & 0.0012 & 0.0086 & 0.0025 & 0.0002 & 0.0001 & 0.0006 & 0.0002 & 0.0017 & 0.0004 & 0.0016 & 0.0009 & 0.0036 & 0.0015 & 0.0154 & 0.0040 \\
SASRec & 0.0030 & 0.0016 & 0.0047 & 0.0021 & 0.0104 & 0.0034 & 0.0012 & 0.0006 & 0.0016 & 0.0008 & 0.0032 & 0.0011 & 0.0091 & 0.0046 & 0.0130 & 0.0059 & 0.0240 & 0.0083 \\
BERT4Rec & 0.0006 & 0.0004 & 0.0011 & 0.0005 & 0.0032 & 0.0010 & 0.0001 & 0.0001 & 0.0001 & 0.0001 & 0.0001 & 0.0001 & 0.0023 & 0.0017 & 0.0036 & 0.0021 & 0.0118 & 0.0039 \\
HGN & 0.0012 & 0.0008 & 0.0023 & 0.0011 & 0.0066 & 0.0020 & 0.0002 & 0.0001 & 0.0010 & 0.0003 & 0.0027 & 0.0007 & 0.0026 & 0.0013 & 0.0058 & 0.0023 & 0.0234 & 0.0062 \\
HSTU & 0.0053 & 0.0037 & \underline{0.0079} & \underline{0.0046} & \underline{0.0140} & \underline{0.0059} & \underline{0.0031} & 0.0021 & 0.0037 & 0.0023 & 0.0052 & 0.0026 & \underline{0.0125} & \underline{0.0081} & \underline{0.0175} & \underline{0.0097} & \underline{0.0279} & \underline{0.0120} \\
ReaRec & \underline{0.0059} & \underline{0.0041} & 0.0068 & 0.0043 & 0.0116 & 0.0054 & 0.0031 & \underline{0.0026} & \underline{0.0038} & \underline{0.0028} & \underline{0.0066} & \underline{0.0034} & 0.0109 & 0.0079 & 0.0130 & 0.0086 & 0.0226 & 0.0107 \\
\midrule
TIGER & 0.0003 & 0.0001 & 0.0005 & 0.0002 & 0.0032 & 0.0007 & 0.0013 & 0.0007 & 0.0019 & 0.0009 & 0.0030 & 0.0012 & 0.0016 & 0.0011 & 0.0028 & 0.0014 & 0.0162 & 0.0042 \\
LETTER & 0.0003 & 0.0002 & 0.0009 & 0.0004 & 0.0036 & 0.0010 & 0.0006 & 0.0004 & 0.0009 & 0.0005 & 0.0036 & 0.0011 & 0.0005 & 0.0003 & 0.0013 & 0.0005 & 0.0136 & 0.0029 \\
LC-Rec & 0.0015 & 0.0007 & 0.0029 & 0.0012 & 0.0079 & 0.0022 & 0.0004 & 0.0002 & 0.0006 & 0.0003 & 0.0015 & 0.0005 & 0.0008 & 0.0005 & 0.0012 & 0.0007 & 0.0048 & 0.0014 \\
SpecGR & 0.0012 & 0.0009 & 0.0020 & 0.0012 & 0.0057 & 0.0020 & 0.0006 & 0.0003 & 0.0012 & 0.0005 & 0.0025 & 0.0008 & 0.0023 & 0.0014 & 0.0040 & 0.0019 & 0.0106 & 0.0033 \\
\midrule
\textbf{DREAM (Ours)} & \textbf{0.0255} & \textbf{0.0222} & \textbf{0.0349} & \textbf{0.0271} & \textbf{0.0819} & \textbf{0.0424} & \textbf{0.0278} & \textbf{0.0207} & \textbf{0.0386} & \textbf{0.0252} & \textbf{0.0827} & \textbf{0.0379} & \textbf{0.0560} & \textbf{0.0483} & \textbf{0.0724} & \textbf{0.0573} & \textbf{0.1239} & \textbf{0.0796} \\
\bottomrule
\end{tabular}}
\end{table*}

Table~\ref{tab:cold} reports the cold-start recommendation
performance, which is the central result of this paper.

(1) DREAM is the best method on every cold-start metric we
measure: all 18 cells across three datasets and six metrics.
The lift over the strongest baseline ranges from $4.3\times$ to
$11.5\times$: cold R@5 rises by $4.3\times$ (Beauty),
$9.0\times$ (Sports), and $4.5\times$ (Toys), while cold N@50 rises by
$7.2\times$, $11.1\times$, and $6.6\times$ on the same datasets, all
relative to the strongest baseline per metric.
The gain is therefore consistent across datasets, metrics, and ranking
depths, confirming that DREAM's progressive refinement pipeline,
rather than dataset-specific tuning, is what drives the improvement.

(2) Static SIDs bottleneck cold-start items.
Among baselines, HSTU and ReaRec---both ID-based sequential
models---consistently provide the second-best cold performance,
whereas SID-based generative baselines rarely stay competitive on
cold-start items.
This pattern indicates that the main difficulty is not autoregressive
recommendation per se, but the quality of the static SID assignment
cold-start items inherit before any user feedback is observed.

(3) SID-based baselines collapse on cold-start items.
TIGER, LETTER, LC-Rec, and SpecGR remain near zero on cold R@10 across
all three datasets.
Even LC-Rec, which injects collaborative signals into backbone
training, does not recover cold-start items, because that enhancement is
applied \emph{after} index construction and cannot repair deficient
SID assignments upstream.
Taken together, the cold table localizes the cold-start bottleneck to
the static SID assignment pipeline, which is precisely where DREAM
intervenes; correspondingly, DREAM's overall gains in
\S\ref{sec:overall_results} are predominantly driven by improvements on
cold-start items.

\subsection{Overall Utility and Warm Trade-off}
\label{sec:overall_results}

\begin{table*}[t]
\centering
\caption{Overall recommendation performance.
\textbf{Bold}: best; \underline{underline}: second best.}
\label{tab:overall}
\resizebox{\textwidth}{!}{
\begin{tabular}{l|cccccc|cccccc|cccccc}
\toprule
\multirow{2}{*}{Method} & \multicolumn{6}{c|}{Beauty} & \multicolumn{6}{c|}{Sports} & \multicolumn{6}{c}{Toys} \\
 & R@5 & N@5 & R@10 & N@10 & R@50 & N@50 & R@5 & N@5 & R@10 & N@10 & R@50 & N@50 & R@5 & N@5 & R@10 & N@10 & R@50 & N@50 \\
\midrule
GRU4Rec & 0.0238 & 0.0146 & 0.0384 & 0.0194 & 0.0989 & 0.0325 & 0.0163 & 0.0102 & 0.0273 & 0.0137 & 0.0752 & 0.0241 & 0.0171 & 0.0106 & 0.0287 & 0.0143 & 0.0828 & 0.0259 \\
SASRec & 0.0163 & 0.0085 & 0.0294 & 0.0127 & 0.0814 & 0.0239 & 0.0111 & 0.0058 & 0.0193 & 0.0085 & 0.0570 & 0.0166 & 0.0221 & 0.0114 & 0.0361 & 0.0159 & 0.0853 & 0.0266 \\
BERT4Rec & 0.0156 & 0.0099 & 0.0280 & 0.0138 & 0.0778 & 0.0246 & 0.0109 & 0.0066 & 0.0174 & 0.0087 & 0.0529 & 0.0163 & 0.0112 & 0.0070 & 0.0190 & 0.0095 & 0.0556 & 0.0173 \\
HGN & 0.0257 & 0.0159 & 0.0442 & 0.0218 & 0.1194 & 0.0382 & 0.0203 & 0.0128 & 0.0337 & 0.0170 & 0.0890 & 0.0290 & 0.0247 & 0.0152 & 0.0426 & 0.0210 & 0.1088 & 0.0354 \\
HSTU & 0.0317 & 0.0217 & 0.0461 & 0.0263 & 0.0999 & 0.0379 & 0.0183 & 0.0122 & 0.0273 & 0.0150 & 0.0646 & 0.0231 & \underline{0.0298} & \underline{0.0203} & 0.0427 & \underline{0.0245} & 0.0874 & 0.0342 \\
ReaRec & 0.0323 & 0.0214 & 0.0470 & 0.0261 & 0.1024 & 0.0381 & 0.0145 & 0.0095 & 0.0234 & 0.0123 & 0.0622 & 0.0207 & 0.0262 & 0.0180 & 0.0379 & 0.0218 & 0.0810 & 0.0311 \\
\midrule
TIGER & \underline{0.0344} & 0.0224 & \textbf{0.0540} & 0.0287 & \underline{0.1305} & \underline{0.0454} & \underline{0.0238} & \underline{0.0152} & 0.0378 & 0.0197 & \underline{0.0959} & 0.0323 & 0.0292 & 0.0180 & 0.0446 & 0.0229 & \underline{0.1144} & 0.0380 \\
LETTER & 0.0316 & 0.0199 & 0.0509 & 0.0261 & 0.1286 & 0.0430 & 0.0235 & 0.0152 & \underline{0.0383} & \underline{0.0199} & 0.0959 & \underline{0.0324} & 0.0271 & 0.0173 & 0.0434 & 0.0225 & 0.1111 & 0.0372 \\
LC-Rec & \textbf{0.0347} & \textbf{0.0228} & \underline{0.0532} & \underline{0.0288} & 0.1255 & 0.0444 & 0.0191 & 0.0121 & 0.0316 & 0.0162 & 0.0832 & 0.0273 & 0.0190 & 0.0120 & 0.0301 & 0.0156 & 0.0723 & 0.0248 \\
SpecGR & 0.0290 & 0.0183 & 0.0494 & 0.0249 & 0.1281 & 0.0420 & 0.0217 & 0.0136 & 0.0347 & 0.0177 & 0.0954 & 0.0308 & 0.0289 & 0.0180 & \underline{0.0476} & 0.0240 & \textbf{0.1225} & \underline{0.0400} \\
\midrule
\textbf{DREAM (Ours)} & 0.0336 & \underline{0.0227} & 0.0531 & \textbf{0.0293} & \textbf{0.1317} & \textbf{0.0475} & \textbf{0.0279} & \textbf{0.0191} & \textbf{0.0443} & \textbf{0.0245} & \textbf{0.1094} & \textbf{0.0392} & \textbf{0.0363} & \textbf{0.0270} & \textbf{0.0527} & \textbf{0.0331} & 0.1123 & \textbf{0.0487} \\
\bottomrule
\end{tabular}}
\end{table*}

Table~\ref{tab:overall} shows that the cold-start gains translate into
competitive overall utility rather than being offset by large
population-level regressions.

(1) Sports is the strongest overall success case.
On Sports, DREAM outperforms all baselines across all six metrics,
with gains of 17.2\% (R@5), 25.7\% (N@5), 15.7\% (R@10), and 23.1\% (N@10)
over the strongest competitor.

(2) Beauty remains competitive overall while benefiting
substantially on cold-start items.
On Beauty, DREAM obtains the best N@10, R@50, and N@50, while staying
within 3.2\% of LC-Rec on R@5 and within 1.7\% of TIGER on R@10.
This pattern is consistent with DREAM's target behavior: large cold-start
improvements lift the overall metrics even though DREAM does not
uniformly dominate every warm-favored metric.

(3) Toys also shows strong overall performance together with
large cold-start gains.
On Toys, DREAM achieves the best result on 5/6 overall metrics and
trails SpecGR only on R@50.
At the same time, it reaches beam-aware cold N@10 $= 0.0573$, showing
that strong cold-start gains are retained together with competitive
overall utility on this dataset.

\begin{table}[t]
\centering
\footnotesize
\setlength{\tabcolsep}{3.5pt}
\caption{Warm-item recommendation performance (R@10 / N@10).
\textbf{Bold}: best; \underline{underline}: second best.}
\label{tab:warm}
\begin{tabular}{l|cc|cc|cc}
\toprule
\multirow{2}{*}{Method} & \multicolumn{2}{c|}{Beauty} & \multicolumn{2}{c|}{Sports} & \multicolumn{2}{c}{Toys} \\
 & R@10 & N@10 & R@10 & N@10 & R@10 & N@10 \\
\midrule
GRU4Rec  & 0.0477 & 0.0241 & 0.0336 & 0.0169 & 0.0365 & 0.0183 \\
SASRec   & 0.0358 & 0.0155 & 0.0235 & 0.0103 & 0.0433 & 0.0191 \\
BERT4Rec & 0.0350 & 0.0173 & 0.0214 & 0.0107 & 0.0239 & 0.0118 \\
HGN      & 0.0551 & 0.0272 & 0.0414 & 0.0210 & 0.0541 & 0.0268 \\
HSTU     & 0.0560 & 0.0319 & 0.0328 & 0.0180 & 0.0507 & 0.0291 \\
ReaRec   & 0.0575 & 0.0318 & 0.0280 & 0.0146 & 0.0457 & 0.0259 \\
\midrule
TIGER  & \textbf{0.0679} & \textbf{0.0361} & \underline{0.0462} & 0.0241 & \underline{0.0577} & \underline{0.0296} \\
LETTER & 0.0638 & 0.0328 & \textbf{0.0471} & \textbf{0.0245} & 0.0565 & 0.0294 \\
LC-Rec & \underline{0.0663} & \underline{0.0359} & 0.0389 & 0.0199 & 0.0391 & 0.0203 \\
SpecGR & 0.0618 & 0.0311 & 0.0426 & 0.0218 & \textbf{0.0612} & \textbf{0.0309} \\
\midrule
\textbf{DREAM (Ours)} & 0.0579 & 0.0299 & 0.0457 & \underline{0.0243} & 0.0465 & 0.0256 \\
\bottomrule
\end{tabular}
\end{table}

Table~\ref{tab:warm} reports warm-item performance.
DREAM does not aim to dominate warm metrics: static-index SID
baselines such as TIGER, LETTER, and SpecGR retain fully optimized
warm-item assignments, whereas CART rewrites the index and
necessarily perturbs some warm SIDs.
The trade-off remains controlled and dataset-dependent.
On Sports, DREAM's warm R@10 ($0.0457$) is within $3.0\%$ of the
strongest warm baseline LETTER ($0.0471$) while its cold R@10 is over
$10\times$ higher than the strongest cold baseline.
On Beauty, DREAM trails on warm R@10 yet lifts cold R@10 from $0.0005$
(TIGER) to $0.0349$.
On Toys, warm metrics still trail the strongest static-index SID
baselines, but DREAM remains best on 5/6 overall metrics while
reaching beam-aware cold N@10 $= 0.0573$.

Overall, these results indicate that DREAM substantially improves
cold-start quality while preserving usable overall and warm-item
performance.

\subsection{Ablation Study}
\label{sec:ablation}

\begin{figure}[t]
\centering
\begin{subfigure}[t]{0.49\columnwidth}
  \centering
  \includegraphics[width=\linewidth]{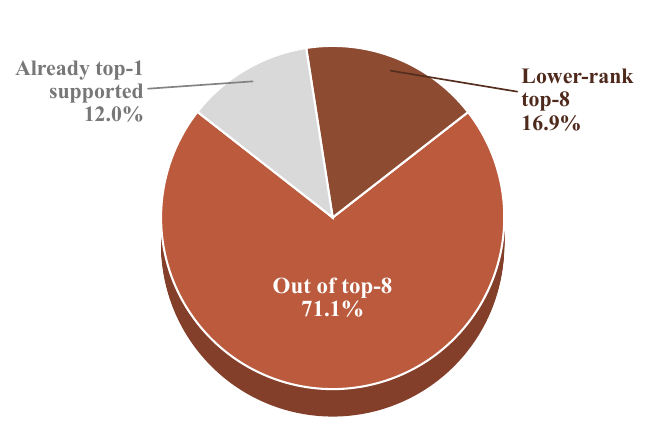}
  \caption{CART prior support.}
  \label{fig:cart_prior_support}
\end{subfigure}%
\hfill
\begin{subfigure}[t]{0.49\columnwidth}
  \centering
  \includegraphics[width=\linewidth]{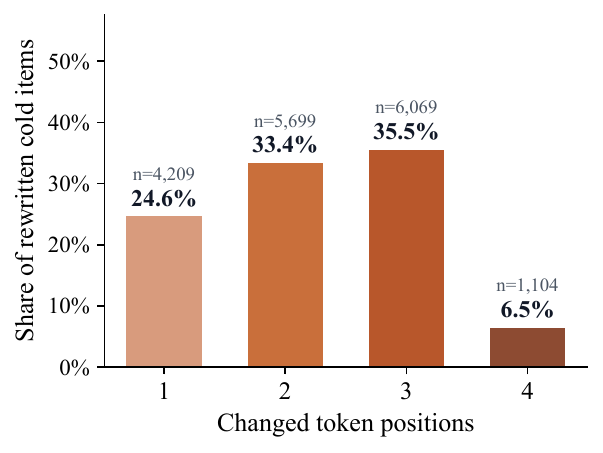}
  \caption{Rewrite locality.}
  \label{fig:cart_rewrite_locality}
\end{subfigure}
\caption{\textbf{Sports CART-side diagnosis.}
CART moves unsupported cold SIDs toward better-supported candidates
with mostly local token edits.}
\Description{Two side-by-side Sports diagnostics. The left plot shows that most inherited static cold-start SIDs are outside CART's top-8 supported pool. The right plot shows that most CART rewrites modify only one to three SID token positions.}
\label{fig:cart_side_diagnosis}
\end{figure}

\begin{table}[t]
\centering
\caption{Ablation study.
Each row removes one or more DREAM stages from the full system.
\emph{Overall} and \emph{Cold} metrics are R@10 / N@10 ($\times 100$).}
\label{tab:ablation}
\resizebox{\columnwidth}{!}{
\begin{tabular}{l|cc|cc|cc}
\toprule
\multirow{2}{*}{Method} & \multicolumn{2}{c|}{Beauty} & \multicolumn{2}{c|}{Sports} & \multicolumn{2}{c}{Toys} \\
& Cold & Overall & Cold & Overall & Cold & Overall \\
\midrule
DREAM
  & \textbf{3.49 / 2.71} & \textbf{5.31 / 2.93} & \textbf{3.86 / 2.52} & \textbf{4.43 / 2.45} & \textbf{7.24 / 5.73} & \textbf{5.27 / 3.31} \\
\quad Compat-Top1
  & 2.17 / 1.31 & 5.04 / 2.65 & 2.60 / 1.56 & 4.20 / 2.27 & 5.13 / 3.16 & 4.77 / 2.70 \\
\quad w/o CPDE
  & 2.19 / 1.31 & 5.02 / 2.63 & 2.53 / 1.53 & 4.19 / 2.26 & 5.05 / 3.15 & 4.72 / 2.69 \\
\quad w/o UC3 \& CPDE
  & 1.93 / 1.18 & 5.04 / 2.65 & 1.98 / 1.20 & 4.09 / 2.22 & 4.35 / 2.73 & 4.82 / 2.68 \\
LC-Rec (Backbone)
  & .29 / .12 & 5.32 / 2.88 & .06 / .03 & 3.16 / 1.62 & .12 / .07 & 3.01 / 1.56 \\
\bottomrule
\end{tabular}}
\end{table}

Table~\ref{tab:ablation} quantifies each stage's contribution.
All Cold and Overall metrics are R@10 / N@10 ($\times 100$).
Removing all DREAM stages (LC-Rec backbone) yields near-zero cold
performance, while adding CART alone (``w/o UC3 \& CPDE'') already
delivers the dominant first repair, with cold N@10 rising by
$\sim\!10\times$, $44\times$, and $22\times$ over the backbone
on Beauty, Sports, and Toys respectively.
UC3 further refines cold quality, and the full DREAM system with
Beam-Aware multi-path decoding achieves the best results on all metrics.
Notably, Compat-Top1 stays close to the ``w/o CPDE'' row across all
datasets, confirming that the Beam-Aware gain is \emph{additive}
multi-path recovery rather than compensation for an upstream regression.

\subsection{Mechanism Diagnosis}
\label{sec:mechanism_diagnosis}

The diagnostics below trace the three compounding factors of early static
commitment, namely unsupported assignment, premature commitment, and the
inference-time single-path constraint, and show how each DREAM stage
resolves the specific factor it targets.

\paragraph{Cold-start failure originates from the static SID commitment,
not from insufficient model capacity.}
On Sports, the static baseline retains visible warm ranking signal
across $N@5/N@10/N@50$ (warm $N@10\!=\!1.99$), while the corresponding
cold NDCG values stay near zero throughout (cold $N@10\!=\!0.03$, a
$\sim 70\times$ gap at the same $K$).
The backbone itself is not globally weak; it is the cold path that
breaks before the backbone has any chance to accumulate behavioral
support.
The system commits each cold-start item to one top-1 SID under content-only
signals, and once that path is registered, the cold-start item is sampled
too rarely during training for the bridge to repair the choice.
The remaining mechanism diagnostics explain how DREAM intervenes at
exactly this commitment point.

\begin{figure*}[!htbp]
\centering
\includegraphics[width=\textwidth]{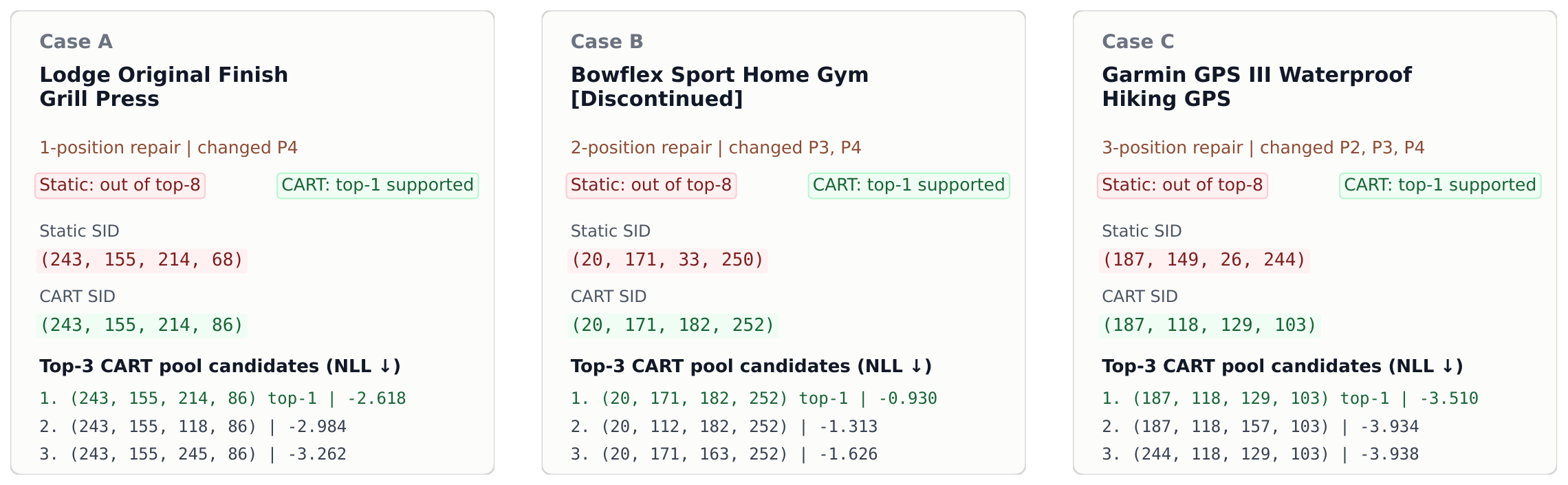}
\caption{\textbf{CART prior-support repair.}
Three deterministic Sports cold-start items whose static SID is absent from
the CART top-8 pool are repaired to the top-1 supported candidate after
1--3 local token edits.}
\Description{Three Sports cold-start case cards. Each card compares the static SID, CART top-1 supported SID, and local token edits for an item whose inherited static SID is absent from the CART top-8 pool.}
\label{fig:cart_case_cards}
\end{figure*}

\paragraph{CART provides the first and largest repair.}
CART delivers the decisive first correction because it rewrites the
assignment space itself rather than only adapting the backbone after the
index is fixed.
CART exports a top-K candidate pool for each cold-start item ($K=8$ here).
We use this pool as a retrospective diagnostic view: the inherited
static SID is \emph{top-8 supported} if it appears anywhere in CART's
exported pool, and \emph{top-1 supported} if it matches the
highest-ranked candidate within the same pool.
This diagnostic asks where the inherited static SID lands inside CART's
exported pool, while CART itself selects the pool's highest-ranked
candidate.
Figure~\ref{fig:cart_prior_support} shows that under this view, the
inherited static SID is outside the top-8 pool for $71.1\%$ of cold
items, only lower-ranked for another $16.9\%$, and already top-1
supported for just $12.0\%$.
Accordingly, CART rewrites $88.0\%$ of cold top-1 SIDs, and among those
rewritten cold-start items, $80.8\%$ move from outside the top-8 pool toward
the pool's top-ranked candidate, while the remaining $19.2\%$ move from
a lower-ranked top-8 candidate toward top-1, with median original rank
$= 3$.
Figure~\ref{fig:cart_rewrite_locality} further shows that this
repair is mostly local rather than arbitrary, since $93.5\%$ of
rewritten cold-start items modify only 1--3 token positions.
Together, these diagnostics indicate that the inherited static SID is
often unsupported, CART rewrites $88.0\%$ of cold-start items, and most
rewritten items are moved from out-of-top8 toward top-1, after which
Sports cold N@10 rises from
$0.03$ to $1.20$ ($43.9\times$).
Figure~\ref{fig:cart_case_cards} makes this repair pattern visually
concrete by showing three Sports cold-start items whose static SID is
absent from the CART top-8 pool, while CART rewrites each item back to
the pool's top-1 supported candidate after only 1--3 local token edits
at the finer (later) SID positions.
The three cards are selected deterministically from the 1-position,
2-position, and 3-position ``out-of-top8 $\rightarrow$ top1'' repair
buckets, illustrating the same aggregate pattern reported above.
These results confirm that the unsupported-assignment factor is the
dominant upstream failure and that CART's prior-support repair resolves
it for the vast majority of cold-start items.

\begin{figure*}[!htbp]
\centering
\begin{subfigure}[t]{0.333\textwidth}
  \centering
  \begin{minipage}[c][0.80\linewidth][c]{\linewidth}\centering
    \includegraphics[width=\linewidth]{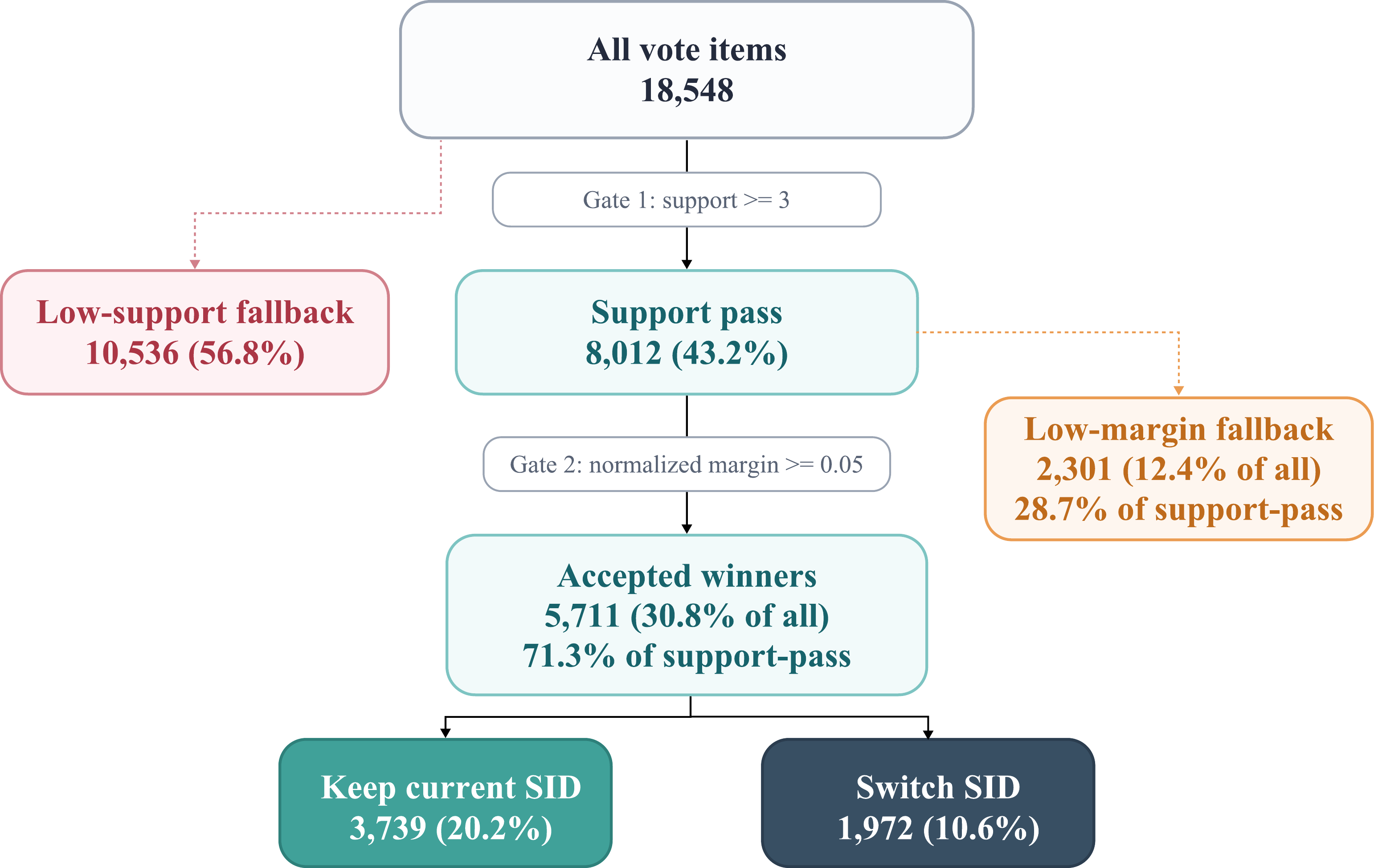}
  \end{minipage}
  \caption{UC3 gate cascade.}
  \label{fig:uc3_gate_cascade}
\end{subfigure}%
\begin{subfigure}[t]{0.333\textwidth}
  \centering
  \begin{minipage}[c][0.80\linewidth][c]{\linewidth}\centering
    \includegraphics[width=\linewidth]{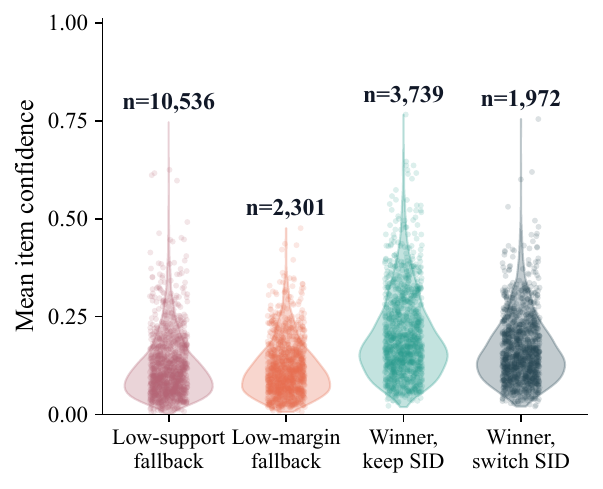}
  \end{minipage}
  \caption{UC3 confidence separation.}
  \label{fig:uc3_confidence_separation}
\end{subfigure}%
\begin{subfigure}[t]{0.333\textwidth}
  \centering
  \begin{minipage}[c][0.80\linewidth][c]{\linewidth}\centering
    \includegraphics[width=\linewidth]{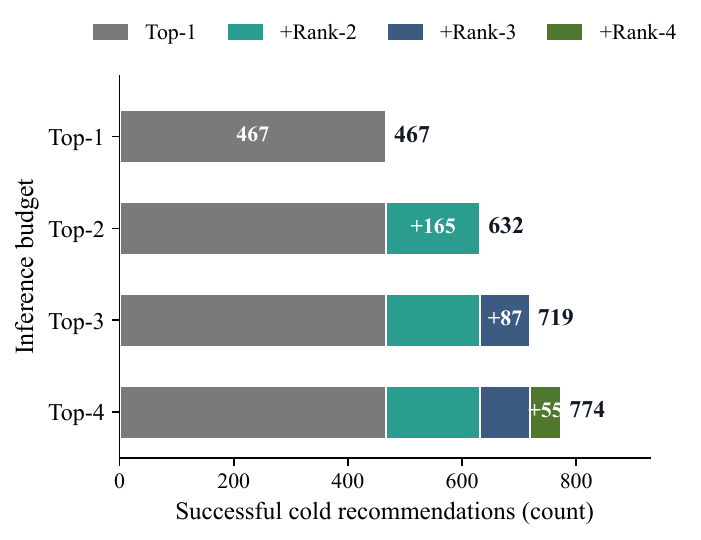}
  \end{minipage}
  \caption{CPDE multi-path recovery.}
  \label{fig:cpde_multi_path_recovery}
\end{subfigure}
\caption{\textbf{UC3 conservative gating and CPDE multi-path recovery
(Sports).}
(a) Of 18{,}548 vote-summary items, $56.8\%$ fall back at the support
gate, $12.4\%$ at the margin gate, and only $10.6\%$ actually switch
SID; (b) winner items exhibit systematically higher mean confidence
than fallback items, supporting the confidence-weighted gating
interpretation; (c) multi-path beam inference rescues $307$ additional
cold hits beyond the top-1 path, lifting cold N@10 from $0.0156$ to
$0.0252$.}
\Description{Three Sports mechanism plots. The first plot shows the UC3 support and margin gate cascade, the second compares confidence for winner and fallback items, and the third shows that CPDE beam inference recovers additional cold hits beyond the top-1 path.}
\label{fig:uc3_cpde_diagnosis}
\end{figure*}

\paragraph{UC3 commits a new SID only when the support is decisive.}
On Sports, $89.4\%$ of cold-start items keep their CART SID, as
Figure~\ref{fig:uc3_gate_cascade} traces this conservative behavior
through a two-gate cascade.
For each cold-start item, UC3 aggregates confidence-weighted votes from
multiple user contexts and accepts a candidate---hereafter a
\emph{winner}, defined as the candidate that survives both the support
and the margin gate---only when the leading candidate is both
sufficiently supported across contexts and decisively ahead of its
runner-up.
Of $18{,}548$ Sports vote-summary items, $10{,}536$ ($56.8\%$) fall
back at the support gate and another $2{,}301$ ($12.4\%$) fall back at
the margin gate; only $5{,}711$ items reach winner status.
Crucially, winner status does not imply an SID change: $3{,}739$
winners reaffirm the current CART SID as the strongest well-supported
option, and a direct diff between the CART and UC3 indices shows that
only $1{,}972$ items ($10.6\%$ of vote items) actually switch away from
the CART SID.
Figure~\ref{fig:uc3_confidence_separation} further shows that winners
have systematically higher mean confidence than fallback items,
consistent with the confidence-weighted interpretation.
UC3 therefore alters a cold SID only when multi-context support is
stable enough to justify an update, and otherwise structurally defers
to the safer CART assignment rather than amplifying weakly supported
rewrites.
This behavior directly addresses the premature-commitment factor:
instead of locking every cold-start item into a single path at the first
opportunity, UC3 gates commitment on sufficient and decisive support,
so that under-supported items remain at the safer prior.


\paragraph{CPDE recovers cold-start items through multi-path inference
within the constrained-decoding trie.}
Beam-Aware registers all surviving beam SIDs in the same prefix trie
used at training time and accepts a generation as a hit when the
produced SID maps to the target through any registered path; no
post-hoc reranker, ensemble, or item-level scoring module is
introduced beyond the standard generative interface.
Within this fixed inference interface, CPDE's gain comes from
preserving recoverable ambiguity rather than from sharpening the
top-1 path: in Table~\ref{tab:ablation}, Compat-Top1 stays close to
UC3 across all three datasets, confirming that the single-path
deployment view is not damaged by CPDE.
On Sports, Figure~\ref{fig:cpde_multi_path_recovery} shows that
multi-path beam inference rescues $307$ successful cold recommendations
beyond the top-1 path, which account for $39.66\%$ of all successful
cold recommendations; the cumulative beam budget reaches $632$ by
Top-2 and $719$ by Top-3, so the gain concentrates in early ranks
rather than in a long tail of accidental matches.
Correspondingly, Sports cold N@10 rises from $1.56$ under Compat-Top1
to $2.52$ under Beam-Aware.
The Sports diagnostics therefore support the intended view of CPDE:
its value lies in preserving alternative valid SID paths so the
decoder can reach a cold-start item through any of them, not in turning
inference into a stronger single-path scorer.
This directly removes the inference-time single-path constraint: cold
items are no longer bound to one registered path, and the decoder can
recover a target through whichever beam path best matches the user
context.


\paragraph{Summary.}
Together, these three diagnostics close the loop on the static
commitment story: each DREAM stage empirically resolves the factor
it was designed to target, and the stage-wise ablation in
Table~\ref{tab:ablation} shows monotone cold-quality gains while
preserving competitive overall utility.


\section{Conclusion}
\label{sec:conclusion}

Cold-start failures in SID-based generative recommendation are not only
representation-learning failures, but also commitment-timing failures:
static tokenizers bind sparsely observed items to a single SID path
before sufficient behavioral support exists, and constrained decoding
then makes poor paths difficult to train and difficult to recover.
DREAM operationalizes this view by making the SID
interface progressively revisable rather than fixed: CART repairs
unsupported assignments with prior-supported candidates, UC3 commits
only under decisive multi-context support, and CPDE keeps residual valid
paths available for multi-path inference. Across three Amazon
benchmarks, DREAM achieves the best result on all 18 cold-start metrics
while preserving competitive overall utility, and the ablation study and mechanism diagnosis align the gains with the intended causal
chain of prior-support repair, conservative commitment, and additive
multi-path recovery. Within the offline item cold-start setting studied
here, these results suggest a broader principle for generative
recommendation: semantic identifiers should be treated as
support-conditioned retrieval interfaces whose reliability can improve
as behavioral evidence accumulates, rather than as immutable item names
fixed before learning begins. This perspective also points to a
natural next step: extending progressive SID refinement to settings
where item evidence, user interests, and catalog composition evolve
continually after deployment.

\clearpage
\section*{GenAI Usage Disclosure}

The authors used ChatGPT to assist with English-language polishing and
clarity-oriented wording revisions. OpenAI Codex was used to debug
experiment-related code and diagnostic scripts. The authors reviewed and
finalized all paper content, experimental results, and claims.

\bibliographystyle{ACM-Reference-Format}
\bibliography{sample-base}

\end{document}